\documentclass[prl,twocolumn,showpacs,superscriptaddress]{revtex4}
\usepackage{amsmath,amssymb,graphicx,epstopdf,color,bm,soul,hyperref}

\begin{document}

\title{Quantum Treatment for Bose-Einstein Condensation in Non-Equilibrium Systems}

\author{H. Flayac}
\affiliation{Institute of Theoretical Physics, Ecole Polytechnique F\'ed\'erale de Lausanne (EPFL), CH-1015 Lausanne, Switzerland}

\author{I. G. Savenko}
\affiliation{COMP Centre of Excellence at the Department of Applied Physics, P.O. Box 11000, FI-00076 Aalto, Finland}
\affiliation{National Research University of Information Technologies,
Mechanics and Optics (ITMO University), Saint-Petersburg 197101, Russia}

\author{M. M\"ott\"onen}
\affiliation{COMP Centre of Excellence at the Department of Applied Physics, P.O. Box 11000, FI-00076 Aalto, Finland}
\affiliation{QCD Labs, Department of Applied Physics, P.O. Box 13500, FI-00076 Aalto, Finland}

\author{T. Ala-Nissila}
\affiliation{COMP Centre of Excellence at the Department of Applied Physics, P.O. Box 11000, FI-00076 Aalto, Finland}
\affiliation{Department of Physics, P.O. Box 1843, Brown University, Providence, Rhode Island 02912-1843, USA}

\begin{abstract}
We develop an approach based on stochastic quantum trajectories for an incoherently pumped system of interacting bosons relaxing their energy in a thermal reservoir. Our approach enables the study of the versatile coherence properties of the system. We apply the model to exciton polaritons in a semiconductor microcavity. Our results demonstrate the onset of macroscopic occupation in the lowest-energy mode accompanied by the establishment of both temporal and spatial coherence. We show that temporal coherence exhibits a transition from a thermal to coherent statistics and the spatial coherence reveals off-diagonal long-range order.
\end{abstract}


\maketitle


\emph{Introduction.---}
A typical signature of Bose--Einstein condensation is the formation of macroscopic occupation in the single-particle ground state of a many-body system in thermal equilibrium. This collective state exhibits distinctive spatial and temporal coherence properties. In solid--state systems, however, the bosons have a finite lifetime and hence we need to study steady-state properties instead of thermal equilibrium~\cite{Wouters2007}. We refer to this scenario as quasi-Bose--Eistein condensation (qBEC). Here, the macroscopic ground-state occupation is induced by the relaxation of the higher excited states which are pumped by some source reservoir. Such qBEC occurs in systems of magnons~\cite{Nikuni1999,Demokritov2007}, indirect excitons~\cite{High2007}, or exciton polaritons~\cite{Kasprzak2006,Balili2007,Lai2007} under nonresonant excitation.

Exciton-polaritons arise from the strong light-matter coupling enhanced in semiconductor microcavities~\cite{kavbamalas}. They behave like bosonic quasiparticles at moderate concentrations ($\le$ $10^{11}$ cm$^{-2}$). Due to their small effective mass polaritons can manifest quantum coherent properties up to room temperatures in wide-bandgap materials~\cite{Christopolous2007, MalpuechZnO2011}.
Polariton qBEC emerges as the result of boson--boson interactions and energy exchange with the environment. In particular, the scattering of polaritons with acoustic phonons~\cite{Tassone1997} plays a key role, as demonstrated in a number of recent experiments~\cite{Wertz2010,Assmann2012}.

The formation of the quasicondensate is associated with emission of coherent laser-like light from the microcavity~\cite{Imamoglu1996, RefChristmann}. However, such emission is not sufficient to prove the existence of qBEC~\cite{ButovKavokin} and since the spatial and temporal coherence properties need to be addressed in more detail. Experimentally, the temporal coherence is described by the second-order temporal coherence function, $g^{(2)}(\tau)$, where $\tau$ is the delay between two photodetection events. In particular, $g^{(2)}(\tau=0)$ exhibits a transition from a thermal $g^{(2)}(0)=2$, to a coherent statistics, $g^{(2)}(0)=1$~\cite{BlochScience,cohbuildKavokin}.
The spatial coherence, also referred to as the off-diagonal long-range order ~\cite{Sarchi2007,PolcohDoan,baas2006,SpatCohPol,Nardin2009}, is characterized by a slowly decaying first order spatial coherence function, $g^{(1)}(\Delta x)$, between regions separated by $\Delta x$.

Whereas experimental techniques based on Michelson interferometry and Hanbury Brown and Twiss (HBT) setups are well established to measure spatial and temporal coherence, a general theory accounting for many-body quantum correlations, environmental interactions, and photon counting is lacking. On one hand, the widely used semiclassical Boltzmann equation approach~\cite{Porras2002, Haug2005, Kasprzak2008, AmoNature,Cao} to the polariton dynamics is based on the assumption of complete incoherence of the system. Thus, the quantum states of the system are taken to be completely uncorrelated.
On the other hand, approaches based on the Gross--Pitaevskii equation~\cite{Carusotto2004} under resonant or nonresonant excitation~\cite{Wouters2007,Keeling2008} are successful in explaining a plethora of recent experiments~\cite{Lagoudakis2011,Christmann2012,Manni2011}, but they assume global coherence and therefore cannot describe phonon-assisted relaxation. The truncated Wigner approches~\cite{Carusotto}, which involve additional noise terms in the Gross--Pitaevskii equation, are based on several limiting assumptions and are not designed to describe multimode systems. Recently, Boltzmann and Gross--Pitaevskii equations have been merged in a classical treatment \cite{Malpuech2014} which, unfortunately, does not provide an accurate description of the onset of coherence.

Master equation approaches allow to account for both coherent and incoherent processes in the polariton dynamics~\cite{SavenkoPRB2011} and spatial coherence has been recently analyzed in a one-dimensional case~\cite{Fisher2014, SavenkoJETP2013}. However, such model, involving a cumbersome hierarchy of coupled and truncated equations, becomes computationally very demanding in higher dimensions, and its application seems to be restricted to 1D structures. 

The model considered in Ref.~\cite{Read2009} includes energy relaxation in a phenomenological way, operate with a classical stochastic field, and require unknown fitting parameters. Thus they seem unable to describe the desired coherence properties.
Furthermore, the long-range interactions prevent the use of powerful quantum methods based on the density matrix renormalization group theory~\cite{DMRG2005}.

In this Letter, we develop a highly parallelizable quantum stochastic approach~\cite{Molmer1} going far beyond the single mode description of Ref.\cite{WoutersQJ} to describe incoherently driven interacting bosons with dissipation caused by a thermal reservoir. The formalism is based on stochastic evolution of the multimode system wave function in its full Hilbert space. It allows the reconstruction of the system density matrix from which correlations such as $g^{(2)}(0)$ or $g^{(1)}(\Delta x)$ can be directly extracted. In addition, we show how to compute the delayed temporal correlation function, $g^{(2)}(\tau)$, from the emission statistics faithfully simulating the Hanbury Brown and Twiss setup. We explicitly apply the method to non-equilibrium polariton condensation.


\emph{The model}.---
We consider a multimode bosonic system with an energy distribution dictated by the dispersion relation $E(\mathbf{k})$ where $\mathbf{k}=(k_x,k_y)$ is the two-dimensional wave vector. The system is in contact with a thermal reservoir at temperature $T$ and driven by an incoherent source with average power $P$. If only the dominant interactions are considered, the system can be described by the Hamiltonian
\begin{eqnarray}
\label{Hamiltonian}
\hat{\cal H}=\hat{\cal H}_{\textrm{kin}}+\hat{\cal H}_{\textrm{p-p}}+\hat{\cal H}_{\textrm{p-ph}}+\hat{\cal H}_{\textrm{pump}}.
\label{EqHamiltonian}
\end{eqnarray}
The first two terms in Eq.~\eqref{Hamiltonian} describe coherent processes: $\hat{\cal H}_{\textrm{kin}} =\sum_\mathbf{k}E_{\mathbf{k}}\hat a^\dagger_{\mathbf{k}}\hat a_{\mathbf{k}}$ is the kinetic energy term, in our case $E_\mathbf{k} = [ {{E_{\textrm{ph}}}\left( \mathbf{k} \right) - \sqrt {{E_\textrm{ph}^2}\left( \mathbf{k} \right) + 4{V^2}} } ]/2$ describes the lower branch of exciton-polariton dispersion. The photonic dispersion is given by $E_\textrm{ph}\left( \mathbf{k} \right) = {\hbar ^2}{k^2}/(2{m_\textrm{ph}})$, the exciton-photon Rabi splitting is $2V$, and we assume infinite exciton mass since $m_\textrm{ex} \gg m_\textrm{ph}$. The creation and annihilation operators of the bosonic  mode with momentum \textbf{k} are denoted by $\hat a_{\textbf{k}}^\dagger$ and $\hat a_{\textbf{k}}$, respectively. The second term
\begin{equation}\label{Hpp}
\hat{\cal H}_{\textrm{p-p}}=\sum_{{\bf k}_1{\bf{ k}}_2\mathbf{p}}U_{\mathbf{k}_1\mathbf{k}_2\mathbf{p}} \hat a_{\mathbf{k}_1}^\dagger \hat a_{\mathbf{k}_2}^\dagger \hat a_{\mathbf{k}_1+\mathbf{p}} \hat a_{\mathbf{k}_2-\mathbf{p}}
\end{equation}
describes the polariton--polariton elastic scattering conserving energy and momentum that corresponds to long-range interaction. The scattering strength, $U_{\mathbf{k}_1\mathbf{k}_2\mathbf{p}}$, is determined by the excitonic fractions, i.e., the Hopfield coefficients, $X_\textbf{k}$, of the initial and final states~\cite{Tassone1999,Piermarocchi}, $U_{\textbf{k}_1\textbf{k}_2\textbf{p}}= U_0 X_{\textbf{k}_1}X_{\textbf{k}_2}X_{\textbf{k}_1 +\textbf{p}}X_{\textbf{k}_2 -\textbf{p}}$, where $U_0=6E_\textrm{b} a_\textrm{B}^2/S$, $E_\textrm{b}$ and $a_\textrm{B}$ are the exciton binding energy and the Bohr radius, and $S$ is the system area.

The last two terms in Eq.~\eqref{Hamiltonian} describe incoherent processes which we treat stochastically. Here, $\hat{\cal H}_\textrm{p-ph}$ accounts for the interaction of the polaritons with a thermal bath of acoustic phonons. To this end, we introduce a Fr\"ohlich-type Hamiltonian~\cite{Tassone1997,Hartwell2010}
\begin{eqnarray}
\label{Hpph}
\hat{\cal H}_\textrm{p-ph}&=& \sum_{\mathbf{k}_1,\mathbf{k}_2} \left[  \int \frac{L_zdq_z}{2\pi}  G_{\mathbf{q}}  \hat a_{\mathbf{k}_1}^\dagger \hat a_{\mathbf{k}_2}\hat b_{\mathbf{q}}+\textrm{h.c.}\right],
\end{eqnarray}
where the sum and integration are performed under the condition of energy and momentum conservation, $|{E_{\textbf{k}_1}-E_{\textbf{k}_2}}|=\hbar\omega_\textbf{q}$, $|\mathbf{q}|^2=|\mathbf{k}_1-\mathbf{k}_2|^2+{q}_z^2$. The phonons are described by the operators $\hat{b}^\dagger_{\mathbf{q}}$ and $\hat{b}_{\mathbf{q}}$, and their dispersion relation, $\hbar\omega_{\mathbf{q}}=\hbar u\sqrt{q_x^2+q_y^2+q_z^2}$, is determined by the speed of sound $u$ of the material \cite{SupplementalMaterial}.
Above, $ G_{\mathbf{q}}$ is the exciton--phonon interaction strength, the microscopic derivation of which and typical values can be found in Refs.~\cite{Hartwell2010}.
We also assume that the phonon reservoir remains thermalized, i.e. $\langle  \hat{b}^\dagger_{\mathbf{q}}\hat{b}_{\mathbf{q}} \rangle=\bar n_\textrm{ph}(\hbar\omega_{\textbf{q}})=\{\exp[\hbar\omega/(k_BT)]-1\}^{-1}$.

The effect of incoherent pumping is described in the rotating-wave approximation by the last term
\begin{equation}
\hat{\cal H}_\textrm{pump} = \hbar\sum_{\mathbf{k}\xi}\left(g_{\mathbf{k}{\xi}} \hat a_{\mathbf{k}}\hat d_{{\xi}}^\dagger+g_{\mathbf{k}{\xi}}^* \hat a_{\mathbf{k}}^\dagger \hat d_{{\xi}}\right),
\label{EqHint}
\end{equation}
of Eq.(\ref{Hamiltonian}), where $\hat{d}_{{\xi}}$ and $\hat{d}_{{\xi}}^\dagger$ are the operators corresponding to the bosonic pumping reservoir in question at temperature $T_\textrm{P}$ and $\langle  \hat{d}^\dagger_{\xi}\hat{d}_{\xi} \rangle=\bar n_\textrm{P}(E_{\textbf{k}})=\{\exp[E_{\textbf{k}}/(k_BT_\textrm{P})]-1\}^{-1}$. The parameters  $g_{\textbf{k} \xi}$ describe the typical linear coupling strengths between the system modes and the reservoir modes. Further, we assume that $g_{\textbf{k} \xi}=\gamma_\mathbf{k}$.
The Hamiltonian~\eqref{EqHint} allows for particle loss which occurs with the rate $\gamma_\mathbf{k}$. The losses coming from~\eqref{EqHint} are mainly caused by leakage of photons from the cavity since their lifetime is much smaller than that of excitons. Thus we can put $\gamma_\mathbf{k}=1/\tau_\mathbf{k}^\textrm{phot}$, where $\tau_{\mathbf{k}}^\textrm{phot}$ is the lifetime of photons.
Without loss of generality, we take into account the natural photon decay from the microcavity using the Hamiltonian (4) (see~\cite{SupplementalMaterial} for details).

Using Eqs.~\eqref{Hpph} and~\eqref{EqHint}, we can derive a Lindblad-type master equation and express the full set of associated quantum jump operators~\cite{Molmer1} as
\begin{eqnarray}
\label{J1}
\hat {\cal J}_{\mathbf{k}}^{+} &=& \sqrt {{\gamma _{\mathbf{k}}}{{\bar n}_\textrm{P}}({E_\mathbf{k}})} {\hat a_{\mathbf{k}}}^\dag,\\
\label{J2}
\hat {\cal J}_{\mathbf{k}}^{-} &=& \sqrt {{\gamma _{\mathbf{k}}}\left[ {{{\bar n}_\textrm{P}}({E_\mathbf{k}}) + 1} \right]} \hat a_{\mathbf{k}},\\
\label{J3}
\hat {\cal J}_{{{\mathbf{k}}_1}{{\mathbf{k}}_2}}^{+} &=&  {\sqrt {\gamma _{{{\mathbf{k}}_1}{{\mathbf{k}}_2}}^\textrm{ph}{{\bar n}_\textrm{ph}}({ {{E_{{{\mathbf{k}}_1}}} - {E_{{{\mathbf{k}}_2}}}} })} } \hat a_{{{\mathbf{k}}_1}}^\dag {\hat a_{{{\mathbf{k}}_2}}},\\
\label{J4}
\hat {\cal J}_{{{\mathbf{k}}_1}{{\mathbf{k}}_2}}^{-} &=&  {\sqrt {\gamma _{{{\mathbf{k}}_1}{{\mathbf{k}}_2}}^\textrm{ph}\left[ {{{\bar n}_\textrm{ph}}({ {{E_{{{\mathbf{k}}_1}}} - {E_{{{\mathbf{k}}_2}}}} }) + 1} \right]} }
{\hat a_{{{\mathbf{k}}_1}}}\hat a_{{{\mathbf{k}}_2}}^\dag,
\end{eqnarray}
where ${{E_{{{\mathbf{k}}_1}}} > {E_{{{\mathbf{k}}_2}}}}$ and we denote the phonon-mediated scattering rate as $\gamma _{{{\mathbf{k}}_1}{{\mathbf{k}}_2}}^\textrm{ph}$ \cite{SupplementalMaterial}.
Equations~\eqref{J1}~and~\eqref{J2} describe the polariton pumping and decay. The average power fed into the polariton system due to interaction with the pumping reservoir is described by $P={\bar n}_\textrm{P}\gamma _{\mathbf{k}}$ .
Equations~\eqref{J3}~and~\eqref{J4} describe transitions between the polariton modes mediated by the phonon reservoir. It should be noted that processes~\eqref{J4} of phonon emission remain even at $T=0$ K.

%
%
%
\begin{figure}[!t]
\includegraphics[width=0.49\textwidth,clip]{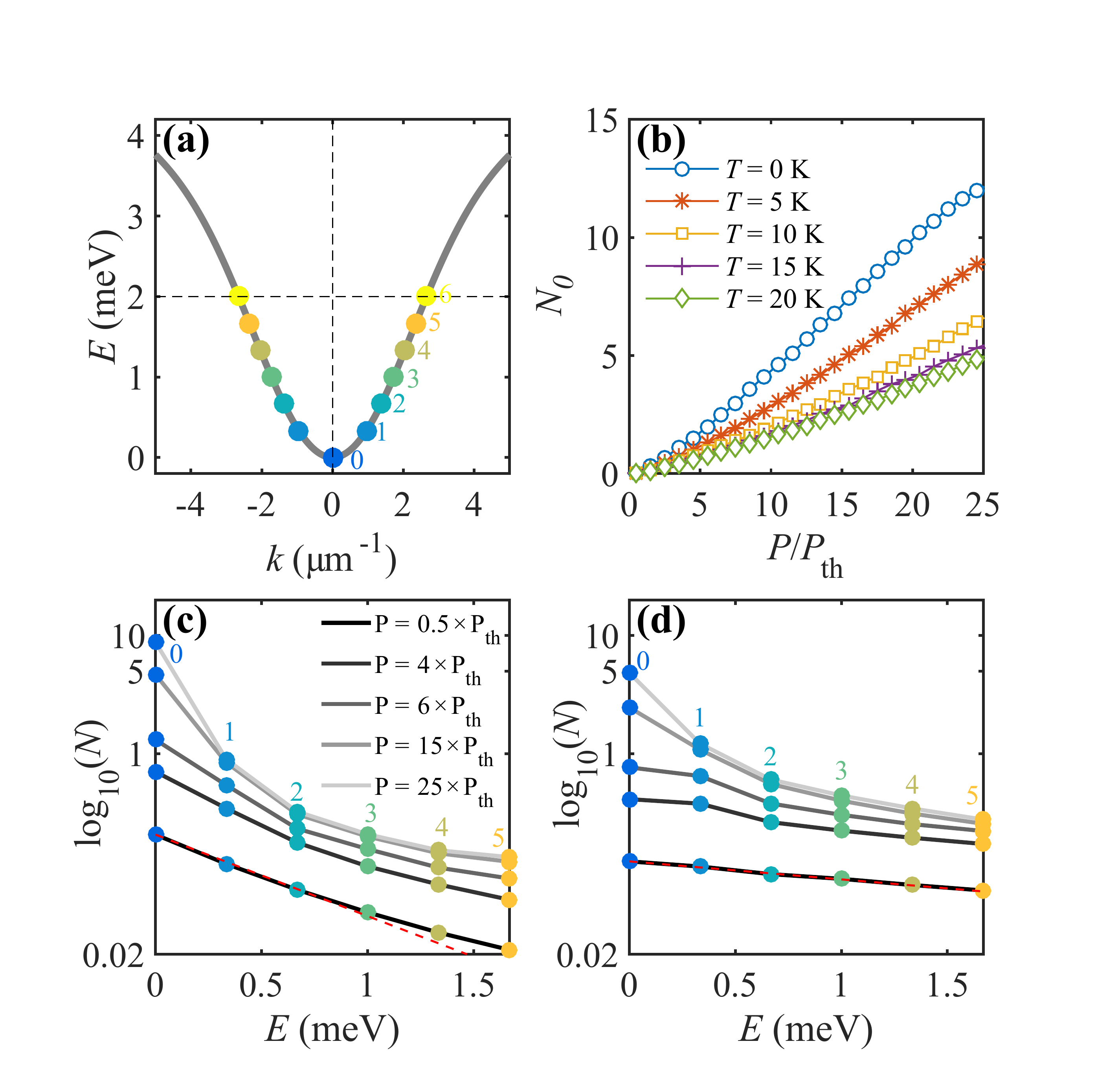}\\
\caption{(color online). (a) Polariton lower dispersion branch (grey line) showing the discrete bosonic modes (dots) used in the computations. (b) Occupation of the lowest mode, $N_0$ as a function of the pump power $P$, for several temperatures in the range $(0-20)$ K (see legend). (c), (d): Occupations of the modes as functions of their energies at (c) $T=5$ K and (d) $T=20$ K for several pump powers in range $(0.5-25)\times P_\textrm{th}$ (see legend). The dashed red lines are the fits by a thermal distribution.}
\label{FigOccupation}
\end{figure}

\emph{The method}.---
The quantum dynamics of the system is simulated using the Monte Carlo wave function technique \cite{Molmer1}. The procedure is based on the evolution of the system wave function through the Schr\"odinger equation,
\begin{equation}\label{Sch}
i\hbar \frac{\partial }{{\partial t}}| \tilde\psi  \rangle={\hat{\tilde{\cal{H}}}}| \tilde\psi  \rangle,
\end{equation}
with the effective non-Hermitian Hamiltonian
\begin{eqnarray}
\label{Heff}
{\hat{\tilde{\cal{H}}}} &=& {\hat{\cal{H}}}
- \frac{{i\hbar }}{2}\sum\limits_\mathbf{k} \hat {\cal J}_{\mathbf{k}}^{+\dagger}\hat {\cal J}_{\mathbf{k}}^{+}
- \frac{{i\hbar }}{2}\sum\limits_\mathbf{k} \hat {\cal J}_{\mathbf{k}}^{-\dagger}\hat {\cal J}_{\mathbf{k}}^{-}
\\
\nonumber
&&- \frac{{i\hbar }}{2}\sum\limits_{\mathbf{k}_1\mathbf{k}_2} \hat {\cal J}_{\mathbf{k}_1\mathbf{k}_2}^{+\dagger}\hat {\cal J}_{\mathbf{k}_1\mathbf{k}_2}^{+}
- \frac{{i\hbar }}{2}\sum\limits_{\mathbf{k}_1\mathbf{k}_2} \hat {\cal J}_{\mathbf{k}_1\mathbf{k}_2}^{-\dagger}\hat {\cal J}_{\mathbf{k}_1\mathbf{k}_2}^{-}
.
\end{eqnarray}
The non-Hermitian part in Eq.~\eqref{Heff} results in an apparent decay of the norm $\langle \tilde\psi(t)|\tilde\psi(t) \rangle$.
We generate a random number, $\eta$, initially and evolve the system by Eq.~(\ref{Sch}). The condition $\langle \tilde\psi(t)|\tilde\psi(t)\rangle\le \eta$ determines if a jump operator occurs or not, see~\cite{Molmer1}. After each jump, we normalize the state $|\tilde\psi(t)\rangle$ again and generate a new number $\eta$.
A single realization, $j$, of this protocol yields a quantum trajectory, $|\tilde\psi(t)\rangle_j$, with $j=1$, $2,\dots,N$. Employing an ensemble of trajectories, we can approximate the system density matrix as
\begin{equation}
\label{EqDensityMatrix}
\hat {\tilde {\rho}} \left( t \right) =\frac{\sum_{j=1}^{N}
{|\tilde\psi( t )} \rangle_{j\,\,j}\langle {\tilde\psi ( t)}|}{N}
\mathop
{\rm{ \rightarrow }}
\limits_{N \to \infty }
\hat{\rho}(t),
\end{equation}
where $\hat { {\rho}} \left( t \right)$ is the actual density matrix of the system. The expectation value of any system observable  $\hat{O}$ can be found from
\begin{equation}
\label{EqObservable}
\langle {\hat O\left( t \right)} \rangle
=
{\rm{Tr}}\left[ { \hat O\hat {{\rho}}(t)} \right]
=
\mathop{\lim}\limits_{N \to \infty }
\left\{
{\rm{Tr}}\left[  { \hat O} \hat{\tilde\rho}(t) \right]
\right\}.
\end{equation}
This method not only allows to significantly reduce the memory consumption by evolving a ket vector instead of a density matrix but it is also ideal for parallelization due to independence of the quantum trajectories. In our computations, we truncate the Hilbert space to a chosen global number of excitations \cite{SavonaArxiv} in addition to the usual truncation per mode, which allows to drastically reduce the dimension of the Hilbert space with negligible loss of accuracy. Due to possible qBEC, the maximum number of excitations in the lowest-energy mode is taken several times larger than for the other states.


\emph{Results and discussion.---}
The parameters we consider correspond to a GaAs-based microcavity having a cylindrical symmetry, with the Rabi splitting $2V = 10$ meV, $m_\textrm{ph}=5\times10^{-5}m_0$, polariton lifetime of $\tau\simeq1/\gamma_k=20$ ps, $E_\textrm{b}=10$ meV, $a_\textrm{B}=10$ nm, and $S=100$ $\mu$m$^2$. The system symmetry and the correlations we compute here allow to consider the radial coordinate $k_r$ only which would not hold anymore if excited states correlation at different angles were under the scope.

Initially the system is prepared in its vacuum state, ${| \tilde{\psi}(0)\rangle }$. Each trajectory is composed of a 500-ps  evolution which is sufficiently longer than the time scales of the processes involved to reach the steady state. For each set of parameters, we average the results over $N=5000$ trajectories.

Figure~\ref{FigOccupation} shows the occupations of the modes, $N_k=\langle {\hat a_k^\dag {{\hat a}_k}} \rangle$ for different pump powers $P$ and temperatures. Our sampling of the dispersion relation is shown in Fig.~\ref{FigOccupation}(a) where we fix the energy step $\Delta E=0.33$ meV between each mode. We approximate the Hopfield coefficients as $X_\mathbf{k}=1/\sqrt{2}$, and hence the polariton--polariton scattering strength is fixed to $U_{\mathbf{k}_1\mathbf{k}_2\mathbf{p}}=U_0/4$. Its value is adjusted to result in a typical chemical potential of $U_0 N_0/4=1$ meV of the lowest-mode if the latter is completely filled and this, to compensate for the low particle number (see outlooks section below). The maximum number of excitations is fixed to $N_0^{\rm{max}}=15$ for the lowest-energy mode and to $N^{\rm{max}}=5$ for the other modes. The polariton--phonon scattering strength is set to $\hbar\gamma_{\mathbf{k}}=\hbar\gamma_0=0.05$ meV.
Finally, we assume that the incoherent pump operator in Eq.~\eqref{J1} is acting only on the highest-energy excited states of the system.
\begin{figure}[!t]
\includegraphics[width=0.49\textwidth,clip]{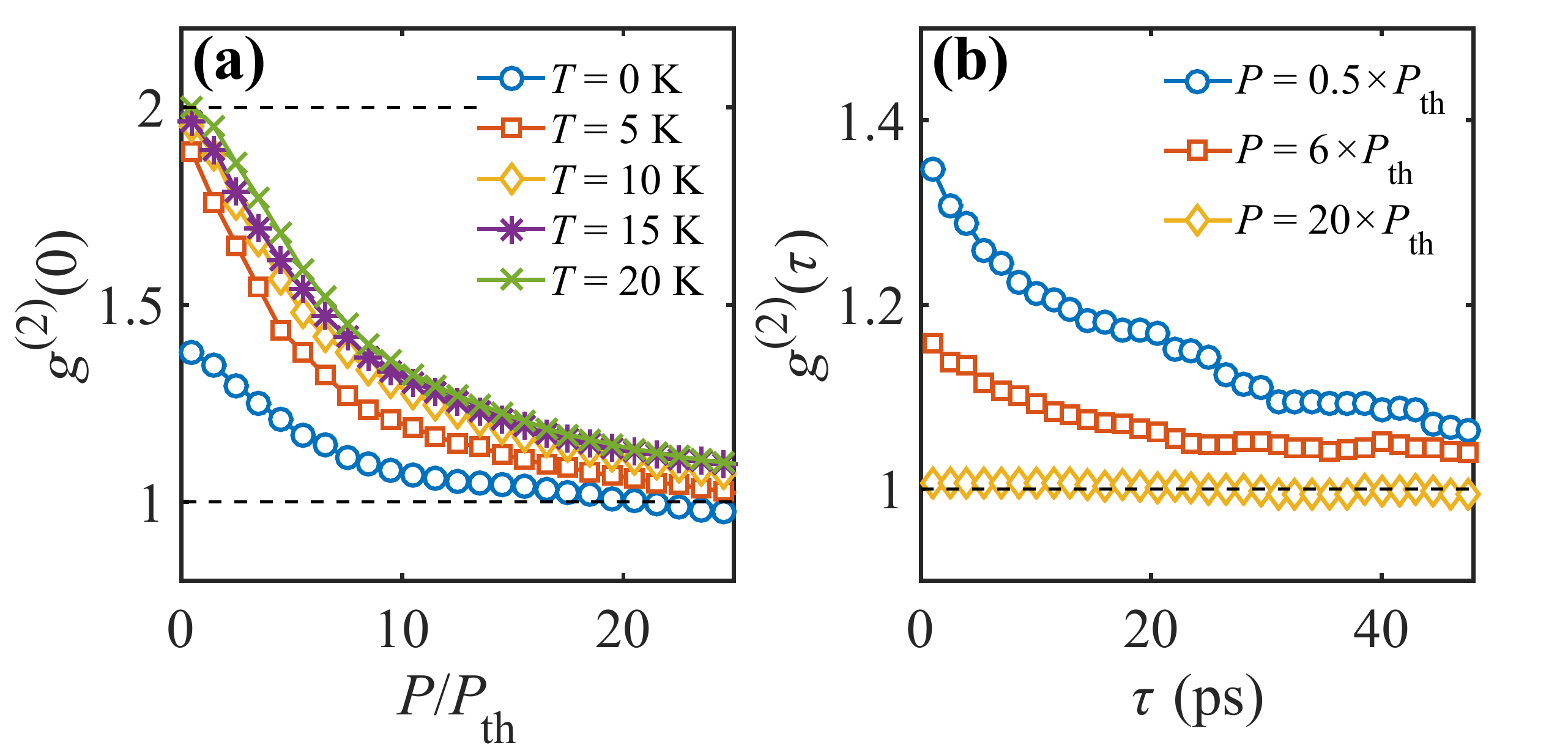}\\
\caption{(color online). (a) Lowest-mode second-order temporal coherence function at zero delay, $g^{(2)}(0)$, as a function of the relative pump power, $P/P_\textrm{th}$, for different phonon temperatures. (c) Finite-delay coherence $g^{(2)}(\tau)$ calculated for 3 different pump powers at $T=0$ K (see legends).}
\label{FigTemporalCoherence}
\end{figure}
Around the threshold power, $P=P_\textrm{th}$, $N_0$ exceeds the population of the other modes. As shown in Fig.~1(b), its value monotonically increases with $P$, faster for lower temperatures, due to the polariton--polariton scattering and phonon-assisted energy relaxation. The highest-energy states in the dispersion are fed by the incoherent pump and play the role of bottleneck modes \cite{Bottleneck}. Thus the latter demonstrate a large occupation even for $P>P_\textrm{th}$ (not shown) although as one can see, the population of the other modes decreases with increasing energy.
The higher the pump power, the greater fraction of particles is observed to reside in the lowest-energy mode. At the lowest investigated pump power, $P=0.5 \times P_{\rm{th}}$, the distribution reads $N_k=N_0\exp[-E_k/(k_BT)]$. Using this we can extract effective polariton temperatures of $\tilde{T}=7.5$ K and $\tilde{T}=23$ K for Figs.~1(c) and~1(d), respectively.

Figure~\ref{FigTemporalCoherence} shows our results for the lowest-mode second-order temporal coherence,
\begin{equation}
\label{g2tau}
{g^{\left( 2 \right)}}\left( \tau  \right)
\mathop {\rm{ = }}
 \frac{{\langle {\hat a_{{0}}^\dag \left( 0 \right)\hat a_{{0}}^\dag \left( \tau  \right){{\hat a}}_{{0}}\left( \tau  \right){{\hat a}}_{{0}}\left( 0 \right)} \rangle }}{{{{\langle {\hat a_{{0}}^\dag \left( 0 \right){{\hat a}}_{{0}}\left( 0 \right)} \rangle }^2}}},
\end{equation}
where the averages are defined using Eq.~\eqref{EqObservable}. From $T=5$ K, we observe a clear crossover from thermal statistics with $g^{(2)}(0)=2$ to a coherent state for which $g^{(2)}(0)=1$ with increasing $P$, as seen in Fig.~\ref{FigTemporalCoherence}(a). With decreasing temperature, the coherence appears at lower pump power as expected.

To compute the delayed $g^{\left( 2 \right)}\left( \tau  \right)$ shown in Fig.~\ref{FigTemporalCoherence}(b), we work on the polariton decay statistics recording the full Eq.~(\ref{J1})-related event history over long 100 ns trajectories. It allows us to build the probability $G^{(2)}(\tau)$ of having two polariton decays within a delay $\tau$. When normalized to the corresponding Poissonian distribution, imposed by the mean steady state occupation, we are able to reconstruct the correlation function and confirm the onset of temporal coherence revealed by $g^{\left( 2 \right)}\left( \tau  \right)\simeq 1$ for $P>P_\textrm{th}$ \cite{FlayacSingle}.

\begin{figure}[ht]
\includegraphics[width=0.49\textwidth,clip]{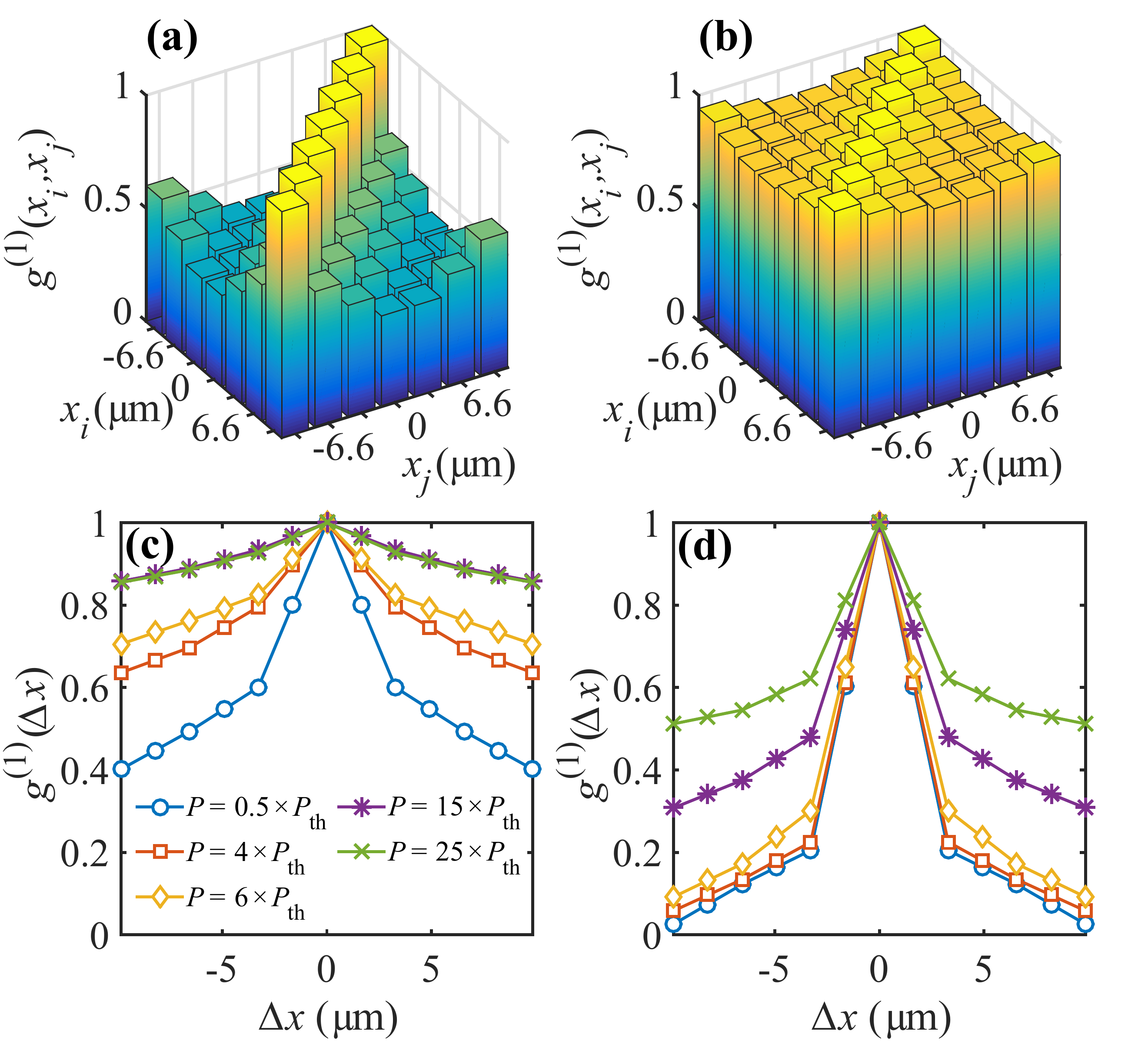}\\
\caption{(color online). Steady-state first-order spatial coherence function $g^{(1)}(x_i,x_j)$ at $T=0$ K (a) below threshold, $P=0.5 \times P_\textrm{th}$, and (b) above threshold, $P=20 \times P_\textrm{th}$, as a function of distance along the sample. Note that periodic boundary conditions are imposed by the Fourier transform. Panels (c) and (d) show $g^{(1)}(\Delta x)=g^{(1)}(\Delta x,0)$ for various pump powers at (c) $T=0$ K and (d) $T=20$ K, where $\Delta x$ is the distance from the centre of the system. }
\label{FigSpatialCoherence}
\end{figure}

Figure~\ref{FigSpatialCoherence} shows the 1D approximation of the first-order spatial coherence function
\begin{equation}
\label{g2x}
g^{\left( 1 \right)}\left( {{x_i},{x_j}} \right)
\mathop {\rm{ = }}
\lim
\limits_{t \rightarrow\infty }
 \frac{{\langle {\hat \psi^\dag \left( {{x_i}},t \right){{\hat \psi}}\left( {{x_j}},t \right)} \rangle }}{{ {\langle {\hat \psi^\dag \left( {{x_i}},t \right)} \rangle \langle {{{\hat \psi}}\left( {{x_j}},t \right)} \rangle } }},
\end{equation}
between two points at positions $x_i$ and $x_j$ in the steady state ($t\rightarrow\infty$).
Here, $\hat\psi(x,t)=\sum_k e^{ikx}\hat{a}_k(t)$. We clearly observe the onset of long-range spatial coherence at large pump powers whereas the coherence decays on short distances for low powers. Comparison of cases $T=0$ K and $T=20$ K, reveals the expected loss of spatial coherence with increasing temperature.
%
%
%


\emph{Outlook.---} In summary, using a stochastic wave function approach, we have analyzed the quantum properties of a non-equilibrium condensate as a function of pump intensity and temperature. Our results exhibit all the characteristic features associated with the Bose--Einstein condensation of such incoherently driven bosonic particles in contact with a phonon bath. To account in future for larger number of bosons and modes, our results can be extended by separating the classical field of each mode, evolving according to Langevin equations, from the quantum fluctuations that would be treated through the quantum jumps approach with the requirement of a very small number of quanta per mode. This will be addressed in a separate study. Finally the impact of decoherence in the form of pure dephasing can be straightforwardly added as a new set of quantum jumps operators \cite{FlayacIO}.

We thank S. Suomela, C. Schneider, J. Pekola and V. Savona for useful discussions. We acknowledge financial support from the Academy of Finland through its Centres of Excellence Program under Grant No. 251748 (COMP) and Grants No. 250280, No. 138903, No. 135794, and No. 272806; the European Research Council under Starting Independent Researcher Grant No. 278117 (SINGLEOUT) and the CCQED EU project; the Government of the Russian Federation, Grant No. 074-U01 and the Dynasty Foundation. The numerical calculations were performed using computer resources of the Aalto University School of Science "Science-IT" project.


\end{document}